\documentstyle[12pt,epsfig]{article} 
\begin{document}
\title{Some Consequences of a Generalization to Heisenberg Algebra in Quantum Electrodynamics}
\author{ A. Camacho 
\thanks{email: acamacho@janaina.uam.mx} \\
Physics Department, \\
Universidad Aut\'onoma Metropolitana-Iztapalapa. \\
P. O. Box 55-534, C. P. 09340, M\'exico, D.F., M\'exico.}
\date{}
\maketitle

\begin{abstract}
In this essay it will be shown that the introduction of a modification to Heisenberg algebra (here this feature means the existence of a minimal obser\-lvable length), as a fundamental part of the quantization process of the electrodynamical field, renders states in which the uncertainties in the two quadrature components vio\-late the usual Heisenberg  uncertainty relation. Hence in this context it may be asserted that any physically realistic generalization of the uncertainty principle must include, not only a minimal observable length, but also a minimal observable momentum.
\end{abstract}
\bigskip
\bigskip

\section{Introduction}

Modifications to Heisenberg algebra have been, since many years, suggested in co\-nnection with quantum gravity, as well as with string theory [1, 2]. Though there are several po\-ssibilities in this context [3] (for instance, we may consider a generalized uncertainty principle that considers only a minimal observable length, one which involves only a minimal observable momentum, or, finally a modification to Heisenberg relation that comprises both cases) there is no sound physical argument that could shed some light upon the correct form that this kind of modifications must have.

In the present essay we will provide an argument that implies that the co\-rrect form must include both, i.e.,  a minimal observable length, but also a minimal observable momentum. This will be achieved considering the effects that a Generalized Uncertainty Principle (GUP) has upon the quantization process of the electromagnetic field. At this point it must be stressed that the main assumption comprises the fact that the classical electromagnetic theory suffers no changes. It will be proved that the effects of GUP appear as a modification that has to be considered in the definition of the so--called creation and annihilation operators. Afterwards, it will be shown that if there is only a minimal observable length (no minimal observable momentum) then the uncertainties in the two quadrature components, of the eigenstates of the occupation number operator [4, 5],  violate the usual Heisenberg relation. 
\bigskip

\section{QED and GUP}
\bigskip

In the case of an electromagnetic field, whose vector potential reads $\vec{A}$, and subject to periodic boundary conditions,  the Fourier decomposition has the following form

{\setlength\arraycolsep{2pt}\begin{eqnarray}
\vec{A}({\vec{x}}, t) = {1\over \sqrt{V}}\sum_{\vec{k}}\sum_{\alpha = 1, 2}\Bigl(c_{\vec{k}\alpha}\hat{e}^{(\alpha)}\exp\Bigl\{i(\vec{k}\cdot\vec{x}- \omega t)\Bigr\} + c^{\ast}_{\vec{k}\alpha}\hat{e}^{(\alpha)}\exp\Bigl\{-i(\vec{k}\cdot\vec{x}- \omega t)\Bigr\}\Bigr).
\end{eqnarray}}

Here the so--called transversality condition [6], a condition that, mathematically, reads $\nabla\cdot\vec{A} = 0$ has been introduced. Additionally, $\hat{e}^{(\alpha)}$ denotes the polarization direction, and $V$ is the volume where the field confined. At this point we also assume that the classical field equations are obtained in the usual way [6]. 

The analogy between the degrees of freedom of the radiation field and a set of uncoupled harmonic oscillators, a fact very known in the quantization process of the electromagnetic field [6], emerges once again. This point is a direct consequence of the assumption that the classical field equations suffer no mo\-difications. 

Hence the Hamiltonian reads

{\setlength\arraycolsep{2pt}\begin{eqnarray}
H = {1\over 2}\sum_{\vec{k}}\sum_{\alpha = 1, 2}\Bigl(\omega^2q^2_{\vec{k}\alpha} + p^2_{\vec{k}\alpha}\Bigr),
\end{eqnarray}}

\noindent where

{\setlength\arraycolsep{2pt}\begin{eqnarray}
q_{\vec{k}\alpha} =  {1\over c}\Bigl(c_{\vec{k}\alpha} +  c^{\ast}_{\vec{k}\alpha}\Bigr),
\end{eqnarray}}

\noindent and

{\setlength\arraycolsep{2pt}\begin{eqnarray}
p_{\vec{k}\alpha} =  -{i\omega\over c}\Bigl(c_{\vec{k}\alpha} -  c^{\ast}_{\vec{k}\alpha}\Bigr).
\end{eqnarray}}
\bigskip

Usually, the quantization is done considering $p_{\vec{k}\alpha}$ and $q_{\vec{k}\alpha}$ as quantum operators, such that $[q_{\vec{k}\alpha}, p_{\vec{k}'\alpha'}] = i\hbar\delta_{\vec{k}\vec{k}'}\delta_{\alpha\alpha'}$ [6]. At this point we suppose that there is a GUP present, the one contains no mimimal observable momentum, only a minimal obser\-vable length [3].

{\setlength\arraycolsep{2pt}\begin{eqnarray}
[q_{\vec{k}\alpha}, p_{\vec{k}'\alpha'}] = i\hbar\delta_{\vec{k}\vec{k}'}\delta_{\alpha\alpha'}\Bigl(\Pi + \beta p^2_{\vec{k}\alpha}\Bigr).
\end{eqnarray}}
\bigskip

Here $\Pi$ denotes the identity operator, while $\beta$ is a constant, which is related to the existence of a minimal observable length [7, 8, 9, 10]. 

At this point one question appears in connection with this GUP, namely, how to define the Fock space? Taking a look at its definition  we see that it depends upon the so--called creation and annihilation operators [4, 5, 6], but in this new case it can be readily seen that the usual definition of creation and annihilation operators (as a function of the position and momentum operators) can not work, since it does not lead to expression (5).

Let us now consider the following possibility, as a generalization for these two operators

{\setlength\arraycolsep{2pt}\begin{eqnarray}
a_{\vec{k}\alpha} = {1\over\sqrt{2\hbar\omega}}\Bigl(\omega q_{\vec{k}\alpha} + i[p_{\vec{k}\alpha} + f(p_{\vec{k}\alpha})]  \Bigr),
\end{eqnarray}}

{\setlength\arraycolsep{2pt}\begin{eqnarray}
a^{\dagger}_{\vec{k}\alpha} = {1\over\sqrt{2\hbar\omega}}\Bigl(\omega q_{\vec{k}\alpha} - i[p_{\vec{k}\alpha} + f(p_{\vec{k}\alpha})]  \Bigr).
\end{eqnarray}}

Here $f(p_{\vec{k}\alpha})$ is a function that satisfies three conditions, namely: (i) in the limit $\beta\rightarrow 0$ we recover the usual definition for the creation and annihilation operators [4, 5, 6], ; (ii) if $\beta \not= 0$, then we have (5), and; (iii) $[a_{\vec{k}\alpha}, a^{\dagger}_{\vec{k}'\alpha'}] = i\hbar\delta_{\vec{k}\vec{k}'}\delta_{\alpha\alpha'}$. It is readily seen that the following function is the only one that satisfies the aforementioned restrictions

{\setlength\arraycolsep{2pt}\begin{eqnarray}
f(p_{\vec{k}\alpha}) = \sum_{n=1}^{\infty}{(-\beta)^n\over 2n+1}p^{2n+1}_{\vec{k}\alpha}.
\end{eqnarray}
\bigskip

Condition (iii) means that the usual results, in relation with the structure of the Fock space, are valid in our case, for instance, the definition of the occupation number operator, $N_{\vec{k}\alpha} = a^{\dagger}_{\vec{k}\alpha}a_{\vec{k}\alpha}$, the interpretation of $a^{\dagger}_{\vec{k}\alpha}$ and $a_{\vec{k}\alpha}$ as creation and annihilation operators, respectively, etc., etc. [4, 5, 6].

We may rewrite the creation and annihilation operators as follows

{\setlength\arraycolsep{2pt}\begin{eqnarray}
a_{\vec{k}\alpha} = {1\over\sqrt{2\hbar\omega}}\Bigl(\omega q_{\vec{k}\alpha} + {i\over\sqrt{\beta}}\arctan(\sqrt{\beta}p_{\vec{k}\alpha})]\Bigr),
\end{eqnarray}}

{\setlength\arraycolsep{2pt}\begin{eqnarray}
a^{\dagger}_{\vec{k}\alpha} = {1\over\sqrt{2\hbar\omega}}\Bigl(\omega q_{\vec{k}\alpha} - {i\over\sqrt{\beta}}\arctan(\sqrt{\beta}p_{\vec{k}\alpha})]\Bigr).
\end{eqnarray}}

Here the operator function $(\sqrt{\beta})^{-1}\arctan(\sqrt{\beta}p_{\vec{k}\alpha})$ denotes the series $p_{\vec{k}\alpha} + \sum_{n=1}^{\infty}{(-\beta)^n\over 2n+1}p^{2n+1}_{\vec{k}\alpha}$. We may also express $q_{\vec{k}\alpha}$ and $p_{\vec{k}\alpha}$ as explicit functions of $a^{\dagger}_{\vec{k}\alpha}$ and $a_{\vec{k}\alpha}$. Indeed, (from now on we omit, for the sake of brevity, all subindices) 

{\setlength\arraycolsep{2pt}\begin{eqnarray}
p = {1\over\sqrt{\beta}}\tan\Bigl(-i\sqrt{{\hbar\omega\beta\over 2}}(a-a^{\dagger})\Bigr),
\end{eqnarray}}
 
{\setlength\arraycolsep{2pt}\begin{eqnarray}
q = \sqrt{{\hbar\over 2\omega}}(a+a^{\dagger}).
\end{eqnarray}}
\bigskip

\section{GUP and Quadrature Components}
\bigskip

Let us now consider a very simple situation, namely, eigenvectors, $\vert n>$, of the occupation number operator and calculate for these states the uncertainties in the gene\-ralized momentum and coordinate variables.

From (11) we have

{\setlength\arraycolsep{2pt}\begin{eqnarray}
p = -i\sqrt{{\hbar\omega\over 2}}\Bigl(a - a^{\dagger}\Bigr)\Bigl[\Pi + {1\over 3}\Bigl(\sqrt{{\hbar\omega\beta\over 2}}(a - a^{\dagger})\Bigr)^2 + {1\over 5}\Bigl(\sqrt{{\hbar\omega\beta\over 2}}(a - a^{\dagger})\Bigr)^4 + ...\Bigr].
\end{eqnarray}}

Therefore we have, up to quadratic order in $\hbar\omega\beta/2$,

{\setlength\arraycolsep{2pt}\begin{eqnarray}
<p>_n = -i\sqrt{{\hbar\omega\over 2}}\Bigl(<n\vert A\vert n> + {1\over 3}{\hbar\omega\beta\over 2}
<n\vert A^2\vert n> + {1\over 5}({\hbar\omega\beta\over 2})^2<n\vert A^4\vert n>\Bigr),
\end{eqnarray}}

{\setlength\arraycolsep{2pt}\begin{eqnarray}
<p^2>_n = -{\hbar\omega\over 2}\Bigl(<n\vert A^2\vert n> + {\hbar\omega\beta\over 3}
<n\vert A^4\vert n> + {23\over 45}({\hbar\omega\beta\over 2})^2<n\vert A^6\vert n>\Bigr).
\end{eqnarray}}

Here we have introduced the definition $A = a - a^{\dagger}$. From the fact that we have eigenstates of the occupation number we deduce that

{\setlength\arraycolsep{2pt}\begin{eqnarray}
<n\vert[a - a^{\dagger}]\vert n> = 0, 
\end{eqnarray}}

{\setlength\arraycolsep{2pt}\begin{eqnarray}
<n\vert[a - a^{\dagger}]^2\vert n> = -(2n+ 1), 
\end{eqnarray}}

{\setlength\arraycolsep{2pt}\begin{eqnarray}
<n\vert[a - a^{\dagger}]^4\vert n> = 3(2n^2 + 2n + 1), 
\end{eqnarray}}

{\setlength\arraycolsep{2pt}\begin{eqnarray}
<n\vert[a - a^{\dagger}]^6\vert n> = -[10n^3 + 19n^2 + 26n + 10]. 
\end{eqnarray}}

From these last expressions we have 

{\setlength\arraycolsep{2pt}\begin{eqnarray}
(\Delta p_n)^2 = \hbar\omega\Bigl[n + 1/2 - \hbar\omega\beta\Bigl(n^2 + n + 1/2\Bigr) \nonumber\\
+ ({\hbar\omega\beta\over 2})^2\Bigl(2.5n^3 + 5.1n^2 + 6.9n+1.5\Bigr) + ...\Bigr],
\end{eqnarray}}

{\setlength\arraycolsep{2pt}\begin{eqnarray}
(\Delta q_n)^2 = {\hbar\over\omega}\Bigl[n + 1/2\Bigr].
\end{eqnarray}}

Hence, it is readily seen that in a power expansion, in terms of $\hbar\omega\beta$, 

{\setlength\arraycolsep{2pt}\begin{eqnarray}
(\Delta p_n)(\Delta q_n) = \hbar\Bigl(n + 1/2\Bigr)\Bigl[1 - \hbar\omega\beta\Bigl({n^2 + n + 1/2\over 2n + 1}\Bigr) \nonumber\\
+ ({\hbar\omega\beta\over 2})^2\Bigl({1.25n^3 + 2.55n^2 + 3.45n+ 0.75\over 2n + 1}\Bigr) + ...\Bigr].
\end{eqnarray}}
\bigskip

Clearly, if we set $\beta =0$, then we recover the usual result [4, 5].
\bigskip

\section{Conclusions}
\bigskip

From the very outset we have restricted the model in such a way that quantum gravity corrections, appear, in the realm of quantum electrodynamics, only as modifications in the Heisenberg algebra for $q_{\vec{k}\alpha}$ and $p_{\vec{k}\alpha}$. 
Since Maxwell equations are the starting point in the canonical quantization procedure [4, 5, 6], a more profound analysis of the effects of a modification to Heisenberg algebra must comprise its consequences upon the classical field equations, an issue that has not been addressed in this essay. Nevertheless, the present approach may be considered as an approximation to the involved physical situation, where the effects of the new Heisenberg algebra, in the classical realm, are neglected.

A new definition, in terms of the position and momentum operators, for the crea\-tion and annihilation operators has been found, such that it allows us to recover the concept of photon. 
Resorting to the eigenvectors of the occupation number operator it was shown that if there is only a minimal observable length, then we may construct states, such that the uncertainties in the two quadrature components violate the usual Heisenberg relation. 

The invariance of (2) under the transformation $q\rightarrow -P/\omega$ and $p\rightarrow Q\omega$, and the fact that the commutation relation for the new variables reads $[Q, P] = i\hbar(\Pi + \gamma Q^2)$, with $\gamma = \beta\omega^2$, allows us to assert that the same conclusion appears if we consider only a minimal observable momentum. Hence, any physically meaningful generalized uncertainty principle must include a minimal observable length and a minimal observable momentum. 
\bigskip
\bigskip
\bigskip

\Large{\bf Acknowledgments.}\normalsize
\bigskip

The author would like to thank A. A. Cuevas--Sosa for his help. 

\end{document}